\documentclass[%
 reprint,
 superscriptaddress,
 amsmath,amssymb,
 aps,
 applied,
]{revtex4-2}
\usepackage{ulem} 
\usepackage{booktabs}
\usepackage{color}
\usepackage{graphicx}% Include figure files
\usepackage{dcolumn}% Align table columns on decimal point
\usepackage{bm}% bold math
\usepackage{ulem}
\usepackage[squaren]{SIunits}

\begin{document}

\title{Nanometer-scale pre-bunched electron beams generated from all-optical plasma-based acceleration
}% Force line breaks with \\

\author{Zhenan Wang}
\thanks{These authors contributed equally to this work.}
\affiliation{State Key Laboratory of Nuclear Physics and Technology, and Key Laboratory of HEDP of the Ministry of Education, CAPT, School of Physics, Peking University, Beijing 100871, China}
 
\author{Zewei Xu}
\thanks{These authors contributed equally to this work.}
\affiliation{State Key Laboratory of Nuclear Physics and Technology, and Key Laboratory of HEDP of the Ministry of Education, CAPT, School of Physics, Peking University, Beijing 100871, China}

\author{Qianyi Ma}
\affiliation{State Key Laboratory of Nuclear Physics and Technology, and Key Laboratory of HEDP of the Ministry of Education, CAPT, School of Physics, Peking University, Beijing 100871, China}

\author{Yuhui Xia}
\affiliation{State Key Laboratory of Nuclear Physics and Technology, and Key Laboratory of HEDP of the Ministry of Education, CAPT, School of Physics, Peking University, Beijing 100871, China}

\author{Letian Liu}
\affiliation{State Key Laboratory of Nuclear Physics and Technology, and Key Laboratory of HEDP of the Ministry of Education, CAPT, School of Physics, Peking University, Beijing 100871, China}

\author{Chenxu Wang}
\affiliation{State Key Laboratory of Nuclear Physics and Technology, and Key Laboratory of HEDP of the Ministry of Education, CAPT, School of Physics, Peking University, Beijing 100871, China}

\author{Thamine Dalichaouch}
\affiliation{Department of Physics and Astronomy, University of California Los Angeles, Los Angeles, California 90095, USA}
\affiliation{Department of Electrical Engineering, University of California Los Angeles, Los Angeles, California 90095, USA}

\author{Xueqing Yan}
\affiliation{State Key Laboratory of Nuclear Physics and Technology, and Key Laboratory of HEDP of the Ministry of Education, CAPT, School of Physics, Peking University, Beijing 100871, China}
\affiliation{Beijing Laser Acceleration Innovation Center, huairou, Beijing 100871, China}
\affiliation{Institute of Guangdong Laser Plasma Technology, Baiyun, Guangzhou 510540, China}

\author{Xinlu Xu}
\email{xuxinlu@pku.edu.cn}
\affiliation{State Key Laboratory of Nuclear Physics and Technology, and Key Laboratory of HEDP of the Ministry of Education, CAPT, School of Physics, Peking University, Beijing 100871, China}
\affiliation{Beijing Laser Acceleration Innovation Center, huairou, Beijing 100871, China}

\author{Warren B. Mori}
\affiliation{Department of Physics and Astronomy, University of California Los Angeles, Los Angeles, California 90095, USA}
\affiliation{Department of Electrical Engineering, University of California Los Angeles, Los Angeles, California 90095, USA}


\date{\today}% It is always \today, today,

\begin{abstract}
High-quality and prebunched electron beams can produce coherent x-rays with high intensity and narrow bandwidth, which is essential for modern light sources. An all-optical scheme based on plasma-based acceleration to produce bright electron beams that are pre-bunched on nanometer-scales is proposed. By using a density modulation created by two low intensity counter-propagating lasers, the phase velocity of the plasma wake excited by an intense driver laser in a uniform plasma can be modulated at a frequency twice that of the colliding lasers and thus turn the injection on and off. The injected electrons are micro-bunched at the Doppler shifted wavelength of the modulated wavelength using the corresponding phase velocity of the gradual expansion of the wakefield. It is demonstrated that by controlling the properties of the drive and colliding lasers, that beams with exotic pre-bunched structures can be produced, which may have critical applications in ultrafast high power x-rays. This extremely compact, all-optical scheme to produce ultra-bright pre-bunched electron beams may therefore enable novel applications for ultrafast x-ray users and arouse general interest in various fields.
\end{abstract}

\maketitle

%\tableofcontents
\section{Introduction}
\label{section1}
High-quality relativistic electron beams with pre-bunched structures have significant applications in modern light sources \cite{gover2019superradiant}. When these pre-bunched electrons radiate, such as when wiggling in magnets, the radiation emitted from different periods adds coherently, greatly enhancing the intensity and narrowing the spectrum \cite{gover2019superradiant}. Furthermore, pre-bunched electron beams can drive free-electron lasers (FEL) \cite{pellegrini2016physics} in a seeded mode to generate fully coherent radiation with stable power in a short undulator, and enable a high radiative energy extraction efficiency when using a tapered undulator. 

Many schemes have been developed within the conventional accelerator community to pre-bunch electrons with millimeter to micron periods, such as modulating the envelope of the laser pulse illuminating a photocathode \cite{li2008nonrelativistic,neumann2009terahertz} or using a mask in a dispersive beamline in combination with an energy chirp \cite{PhysRevLett.101.054801} or in front of a longitudinal to transverse exchange region \cite{PhysRevLett.105.234801}. Beam space-charge oscillations have also been explored to enhance the bunching rather than diminish it \cite{PhysRevLett.106.184801, PhysRevLett.116.184801, liang2023widely}. To produce nanometer-scale bunched beams for x-ray generation, several concepts have been proposed using lasers to manipulate the electron phase space in combination with magnetic chicanes and undulators. Such concepts include cascaded high-gain harmonic generation (HGHG) \cite{yu2000high, allaria2013two} and echo-enabled harmonic generation (EEHG) \cite{stupakov2009using, xiang2010demonstration, zhao2012first, ribivc2019coherent}. These concepts essentially imprint the field modulation of a laser pulse onto the beam current profile with a harmonic number of $\sim$100.

Plasma-based accelerators (PBA) \cite{PhysRevLett.43.267, chen1985acceleration, RevModPhys.81.1229} can sustain $\giga\volt/\centi\meter$ acceleration gradients in centimeter-long plasmas and thus hold the promise of building compact and economical XFELs \cite{wang2021free, pompili2022free, PhysRevLett.129.234801, labat2023seeded, vh62-gz1p}, which can serve a significantly broader x-ray user community. PBA might also enhance the capability of existing facilities. Ultra-short and ultra-intense laser pulses or charged particle beam drivers excite nonlinear plasma wave wakes that can become large enough to trap (self-inject) plasma electrons through various mechanisms \cite{PhysRevE.58.R5257, chen2006electron, PhysRevLett.104.025003, PhysRevLett.100.215004, kalmykov2009electron, kostyukov2009electron, PhysRevLett.112.035003}. The plasma wakes are nonlinear oscillations that are phased such that the phase velocity is near the speed of light. The electrons are usually self-injected into the first wavelength (oscillation) of the wakefield.

The resulting beams typically exhibit trivial current profiles devoid of microstructures. However, a pre-bunched beam with a period of 10s of microns can be formed if electrons are trapped in a series of wakes \cite{lundh2013experimental}. Ionization injection may produce beams with $\sim100~\nano\meter$ period \cite{xu2016nanoscale} due to the phase-dependent tunneling ionization probability \cite{ADKionizationmodel1986}. Higher-dimensional effects seem to limit the modulation period of these self-injected beam. Recently, we showed through simulations that the combination of a density downramp and a periodic density modulation can produce bright electron beams with nanometer period \cite{xu2022generation}. A GeV-class electron beam driver propagating in the downramp produces a wakefield with a wavelength of the first oscillation that gradually increases, thereby reducing the phase velocity below a trapping threshold. The additional modulation lead to an periodic oscillation to the phase velocity, which turns the injection on and off. Evidence of these pre-bunched beams from PBA have been found in experiments \cite{PhysRevLett.125.014801, laberge2024revealing}.

\begin{figure*}
\centering
\includegraphics[width=0.7\linewidth]{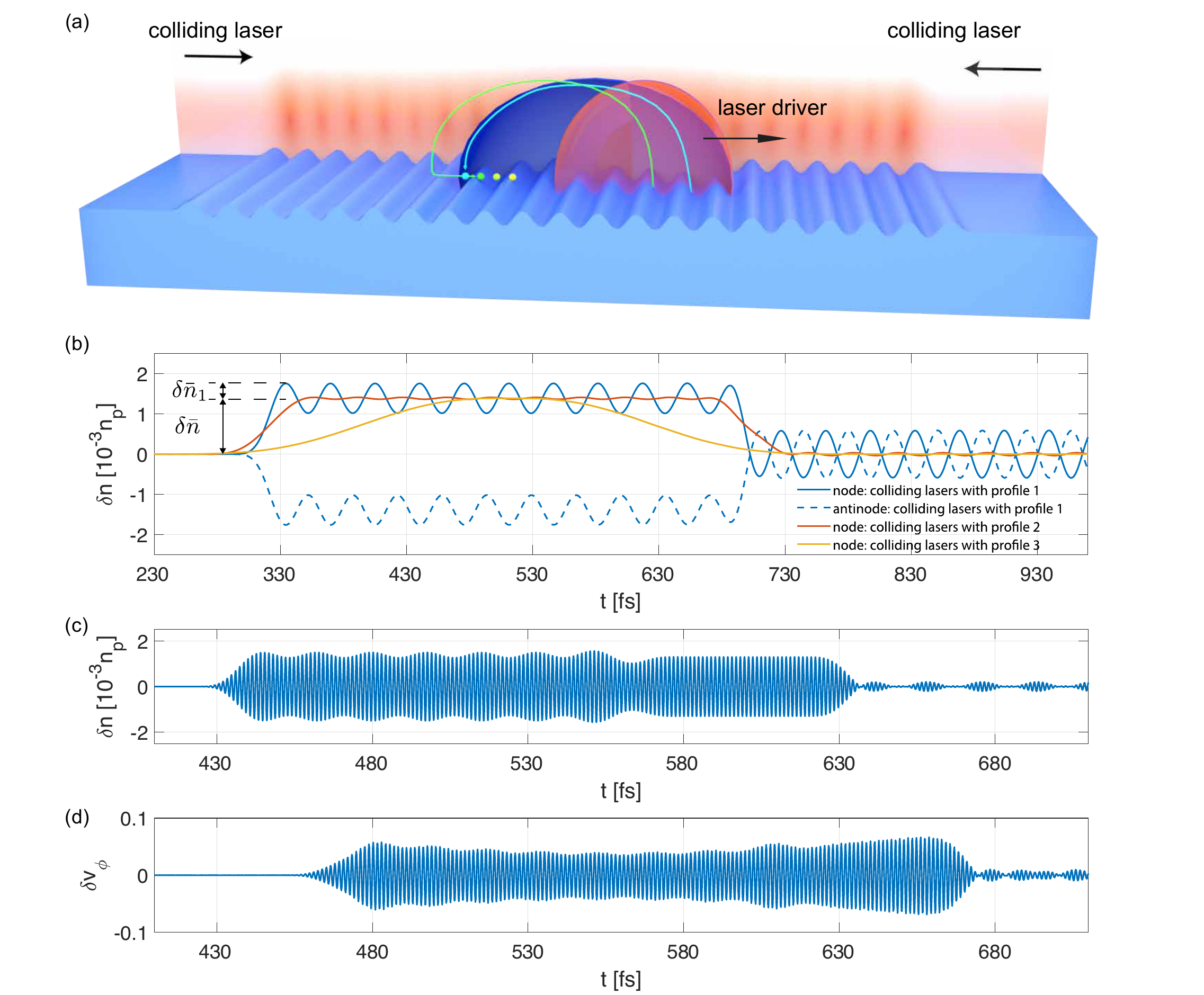}
\caption{Schematic for all-optical pre-bunched beam generation. (a) Two long and low intensity colliding lasers modulate the plasma density through the ponderomotive force when they superimpose, which can turn the injection on and off frequently in a nonlinear wake driven by an intense drive laser to form a pre-bunched electron beam (not to scale). (b) The density perturbation at $z=84.2~\micro\meter$ (solid lines) and $84.4~\micro\meter$ (dashed line). Profile 1: $t_\mathrm{rise}=t_\mathrm{fall}=56~\femto\second$, $t_\mathrm{flat}=337~\femto\second$, profile 2: $t_\mathrm{rise}=t_\mathrm{fall}=112~\femto\second$, $t_\mathrm{flat}=293~\femto\second$ and profile 3: $t_\mathrm{rise}=t_\mathrm{fall}=309~\femto\second$, $t_\mathrm{flat}=0$. (c) The density perturbation encountered by the head of a drive laser which is 280 fs behind the colliding laser. (d) The phase velocity modulation of the tail of the wake excited by a drive laser with $a_0=0.5$.} \label{schematic_plot}
\end{figure*}

In this article, we propose an all-optical (three laser pulses) method to generate high-quality pre-bunched electron beams with period of a few nanometers in a nonlinear laser-driven plasma wakefield. Two low-intensity, hundreds of fs long counter-propagating and co-polarized laser pulses collide inside a uniform underdense plasma and create a sinusoidal electron density modulation through the ponderomotive force where they overlap \cite{sheng2003plasma}. The use of a uniform plasma instead of a plasma with a density downramp makes the set up simpler. A third intense and short laser pulse driver co-propagates and is delayed with respect to the front of the co-propagating low-intensity laser. It is polarized in the orthogonal plane with respect to the low intensity pulses. This intense pulse creates a fully nonlinear plasma wave wakefield and drives self-injection as the wakefield expands slowly through the evolution of the driver laser. The counter-propagating low intensity pulse overlaps with the drive and the other low intensity pulse (for some duration) leading to the modulation of the electron density. The injected electrons are then mapped into a pre-bunched beam with a significantly Doppler shift of the modulated wavelength \cite{xu2017high, xu2022generation}. 

The parameters of the low intensity colliding lasers (e.g., pulse profile, polarization and frequency) can be utilized to control the characteristics of the modulated plasma density encountered by the driver to potentially produce pre-bunched beams with exotic features, such as chirped periods or slice dependent modulation amplitudes, which can generate x-rays with exotic properties. This all-optical system is found to generate ultra-bright and nanometer-scale pre-bunched beams, while offering the advantages of compactness, simplicity, and synchronization, providing a feasible path to the construction of compact high power, ultrafast and fully coherent x-ray sources.

\section{Modulation of the phase velocity of a wake using two counter-propagating low intensity laser pulses\label{section2}}
The concept is illustrated in Fig. \ref{schematic_plot}(a), where two identical and counter-propagating 800 nm wavelength, linearly polarized (LP) laser pulses with spot sizes of $10~\micro\meter$ and peak normalized vector potentials of $a_\mathrm{1}=\frac{eE_\mathrm{1}}{mc\omega_\mathrm{1}}=0.002$ propagate in a long and uniform plasma with $n_\mathrm{p}=10^{19}~\centi\meter^{-3}$. For simplicity, the frequency and the peak electric field of the colliding lasers, $\omega_\mathrm{1}$ and $E_\mathrm{1}$ respectively, are the same, and $\omega_\mathrm{p}=\sqrt{\frac{n_\mathrm{p}e^2}{m\epsilon_0}}$ is the plasma frequency, $m$ and $e$ are the electron mass and charge, $c$ is the speed of light in vacuum, and $\epsilon_0$ is the vacuum permittivity. Each colliding laser has a temporal profile consisting of rising and falling edges of $t_\mathrm{rise}=t_\mathrm{fall}=10\omega_\mathrm{p}^{-1}~(56~\femto\second)$, and a flattop of duration $t_\mathrm{flat}=60\omega_\mathrm{p}^{-1}~(337~\femto\second)$. The profile of the rising and falling edge can be found in Methods. The plasma exists between $z=0$ and $100c/\omega_\mathrm{p}~ (168~\micro\meter)$. The heads of the colliding lasers simultaneously reach the left and right plasma boundaries at $t=0~\femto\second$, respectively. When counter-propagating lasers overlap inside the plasma, a standing wave is formed, and its ponderomotive force pushes the electrons from the peaks to the troughs of the force while heavy ions remain at rest for the duration of overlapping of the lasers \cite{sheng2003plasma}. After an initial rise, the density perturbation in the laser overlapping region reaches a quasi-steady state since the charge-separation force between the electrons and ions balances the ponderomotive force. One-dimensional (1D) analysis gives the expression of the spatial density perturbation as \cite{xu2022generation} $\delta n\approx \delta \bar{n} \mathrm{cos}(2k_\mathrm{1} z)$, which is a sinusoidal modulation at $2k_\mathrm{1}$ with an amplitude of $\delta \bar{n} = 2  a_\mathrm{1}^2\left(\frac{\omega_\mathrm{1}}{\omega_\mathrm{p}}\right)^2 n_\mathrm{p}$. 

One-dimensional simulations are performed using the fully relativistic and nonlinear PIC code OSIRIS \cite{fonseca2002high} to study the plasma response to two counter-propagating lasers, and the parameters can be found in the Methods section. Fig. \ref{schematic_plot}(b) shows the density evolution at a node (solid line) and an antinode (dashed line) of the electric field. The density perturbation during the quasi-steady state agrees well with the theoretical prediction $\delta \bar{n}\approx 0.0014n_\mathrm{p}$. In addition, part of the density perturbation oscillates around $\delta \bar{n}$ with an amplitude of $\delta \bar{n}_1$ with a frequency of $\omega_\mathrm{p}$ after the initial rise. The value of $\delta \bar{n}_1$ can be diminished by using lasers with larger $t_\mathrm{rise}$, as demonstrated by the red and yellow lines in Fig. \ref{schematic_plot}(b). This quasi-steady state exists during the time the lasers overlap, which is $\sim350$ fs for these two points. After the lasers pass through each other, the ponderomotive force disappears. If the lasers decay within a few plasma periods, the spatial density modulation continues to oscillate at the plasma frequency due to the ions' restoring force as shown in Fig. \ref{schematic_plot}(b). The amplitude of these oscillations can be controlled through the fall time ($t_\mathrm{fall}$) of the lasers. As these oscillations are at the plasma frequency, they do not lead to modulations of the wake's phase velocity.

\begin{figure}[ht]
\centering
\includegraphics[width=1\linewidth]{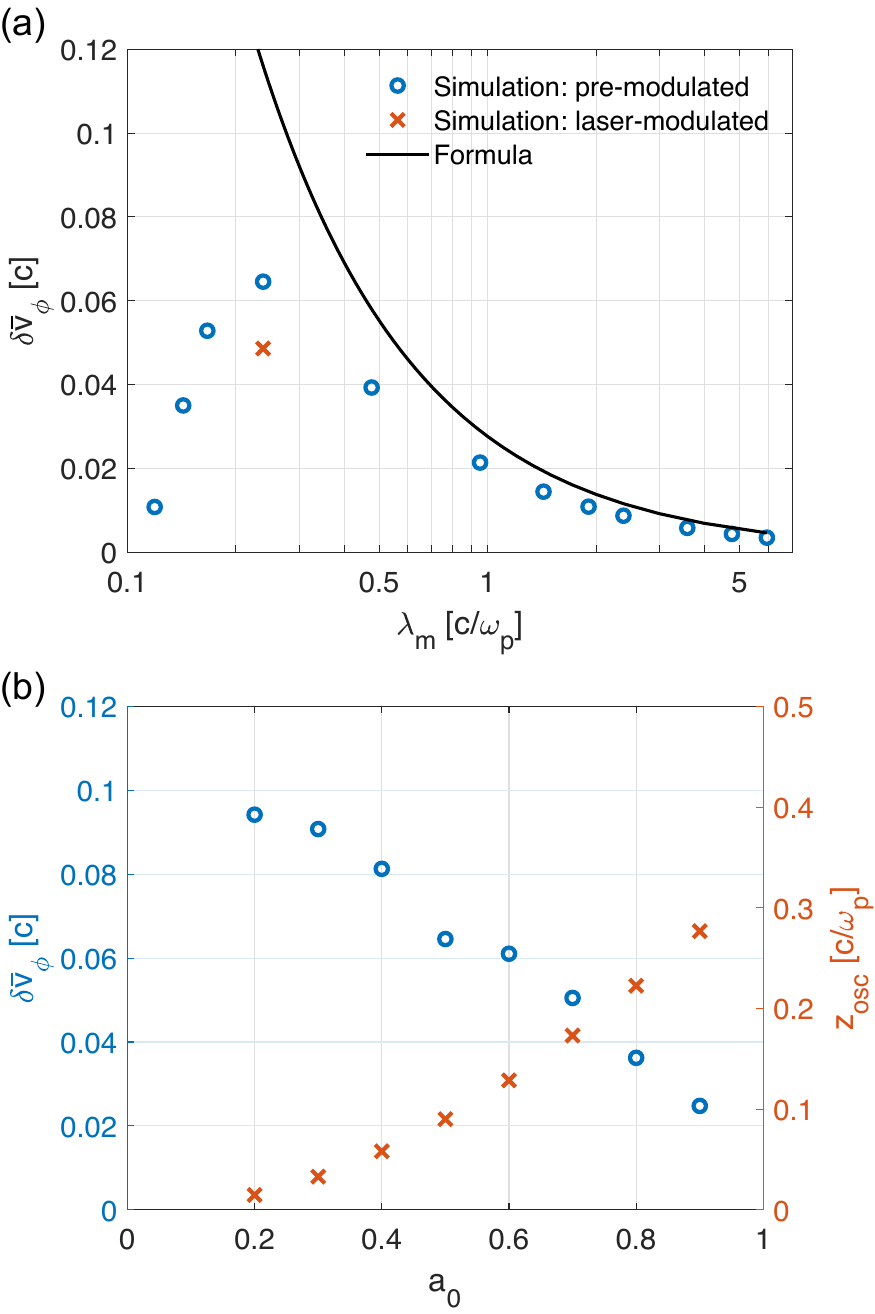}
\caption{The amplitude of the phase velocity modulation extracted from 1D PIC simulations with different modulation periods $\lambda_m$ (a) and different $a_0$ (b). The result predicted by Eq. \ref{deltavphi} is in black and the simulation results are in blue and red. The excursion of the plasma electrons under different $a_0$ is shown in (b) by the red crosses.} \label{phase_velocity}
\end{figure}

\begin{figure*}
\centering
\includegraphics[width=0.7\linewidth]{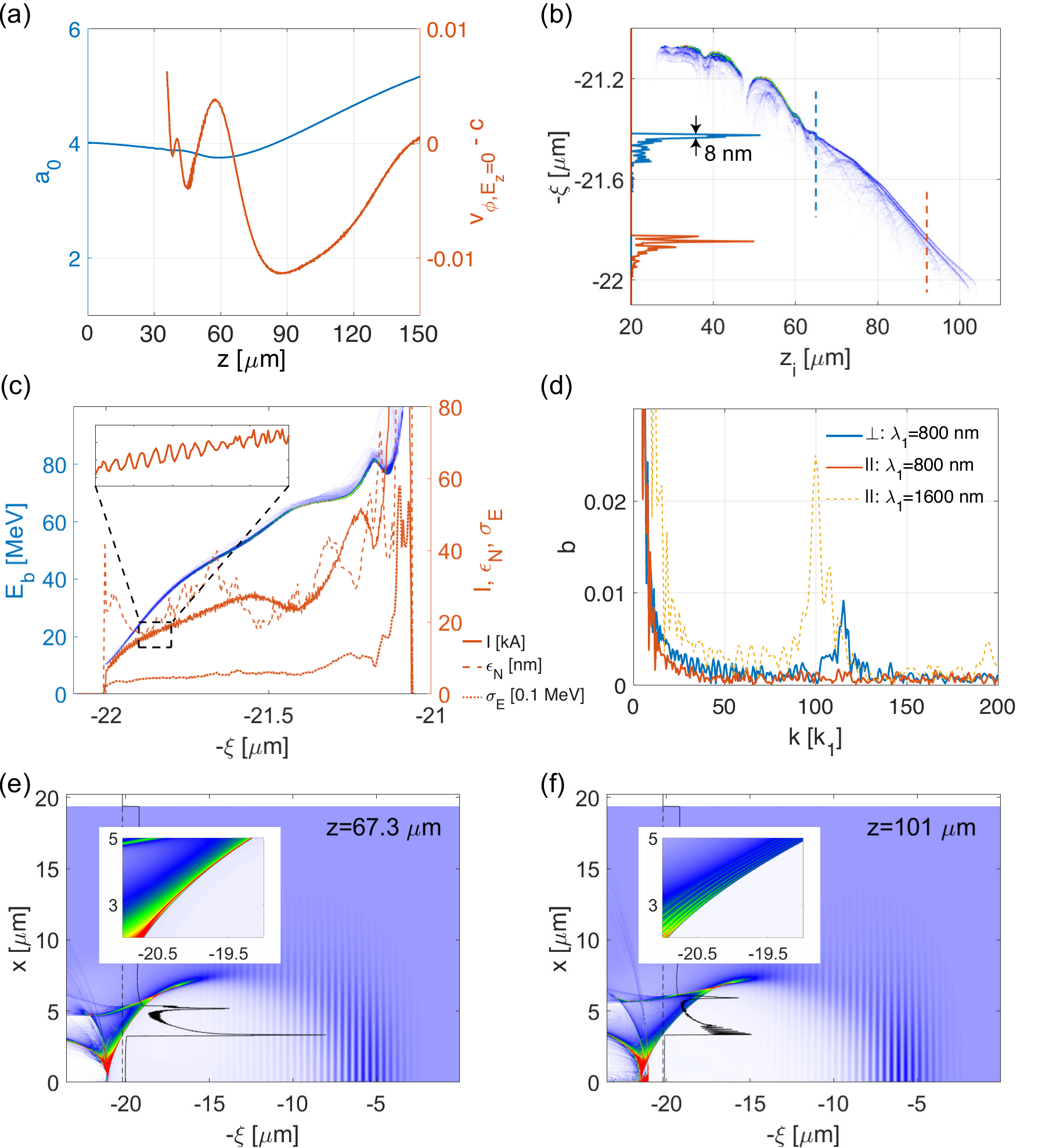}
\caption{All-optical generation of pre-bunched beams. (a) The evolution of the peak $a_0$ of the drive laser and the velocity of the wake center. (b) The charge density distribution of the injected electrons in the ($z_\mathrm{i}, \xi$) space. The blue and red lines show the $\xi$-distribution at $z_\mathrm{i}=65$ and 92 $\micro\meter$. (c) The longitudinal phase space and the current profile of the injected electrons at $t=912~\femto\second$. (d) The bunching factor for electrons with $\xi<21.4~\micro\meter$ when the polarization of the drive laser is perpendicular (blue) or parallel (red) to that of the colliding lasers. A result corresponding to the case where the drive laser is parallel to the colliding lasers and $\lambda_\mathrm{1}=1600~\nano\meter$ is shown in yellow. The plasma electron density distribution at $z=67.3~\micro\meter$ (e) and $101~\micro\meter$ (f). The insets show the enlarged density distribution near the wake tail. The black solid lines represent the density lineout at $\xi=20.2~\micro\meter$, marked by the dashed lines. }\label{injection}
\end{figure*}

When an ultra-short and ultra-intense laser or charged particle beam driver propagates through a uniform plasma, its ponderomotive force or space-charge force pushes the electrons forward and sideways from their equilibrium positions, thereby exciting a plasma wave wake with a phase velocity approximately equal to the group velocity of the driver ($v_\mathrm{d}\sim c$) \cite{RevModPhys.81.1229}. The plasma oscillation frequency depends on the local plasma density as $\omega_\mathrm{p}(z)\propto \sqrt{ n_\mathrm{p}(z) }$ (nonlinear corrections can be included if necessary) , therefore the phase velocity of the wake in a nonuniform plasma can deviate from $v_\mathrm{d}$ \cite{PhysRevA.33.2056, PhysRevLett.119.064801}. If the characteristic scale length of the plasma density modulation is much longer than the plasma wavelength $\lambda_\mathrm{p}$, all electrons within the wavelength of the wakefield experience an approximately constant ion density during their oscillations and we have\cite{xu2017high} $v_{\phi}\approx \frac{v_\mathrm{d}}{1-(\mathrm{d}\omega_\mathrm{p}/\mathrm{d}z)\omega_\mathrm{p}^{-1}\xi}$, where $\xi\equiv ct-z$ is the coordinate in the speed of light frame. A small-amplitude density modulation of the form $\delta\bar{n}\mathrm{cos}(k_\mathrm{m}z)$ can thus induce a phase velocity modulation given by 
\begin{align}
    \delta v_\phi \approx \delta \bar{v}_\phi\mathrm{sin}(k_\mathrm{m}z) = -c\frac{\delta \bar{n}}{n_\mathrm{p}}  \frac{k_\mathrm{m}}{2}\xi \mathrm{sin}(k_\mathrm{m}z).\label{deltavphi} 
\end{align}
The phase velocity modulation amplitude is proportional to the density modulation amplitude, the modulation wavenumber, and the distance behind the driver. 

In the proposed pre-bunching scheme, we are interested in a density modulation whose period is much smaller than the plasma wavelength, i.e., $\lambda_\mathrm{m} \ll \lambda_\mathrm{p}$. The excursion of the oscillating electrons $z_\mathrm{osc}$ is typically much longer than the modulation wavelength, thus they experience a varying ion density during one oscillation cycle. Their oscillation frequency, therefore, is not solely determined by the ion density at the equilibrium position but rather by an integrated contribution from the ions they encounter. As a result, the corresponding $\delta \bar{v}_\phi$ is smaller than that predicated by Eq. \ref{deltavphi} \cite{xu2022generation}. 

In these 1D simulations, an 800 nm wavelength laser pulse with $a_0=0.5$ and a full-width-half-maximum (FWHM) pulse duration of 21 fs is used to excite the plasma wakefield. The head of the laser pulse is 280 fs behind the head of the co-propagating colliding laser. Fig. \ref{schematic_plot}(c) shows simulation data of the plasma density encountered by the head of the drive laser, which is a sinusoidal modulation with $2\omega_\mathrm{1}$. Colliding lasers with slow rising edges can suppress the envelop oscillation. The variation in phase velocity of the tail of the first wake caused by this density modulation is shown in Fig. \ref{schematic_plot}(d). The modulation is approximately sinusoidal with an amplitude of $\sim0.05c$, which is less than the predication of Eq. \ref{deltavphi} ($0.12 c$ for $\xi=2\pi c/\omega_\mathrm{p}$). This is because the excursion of plasma electrons is $0.1c/\omega_\mathrm{p}$, which is comparable with the modulation period $\lambda_1/2\approx 0.24c/\omega_\mathrm{p}$. 

To quantify the degradation to the modulation of $v_\phi$, we scan $\delta \bar{v}_\phi$ v.s. $\lambda_\mathrm{m}$ with fixed $\delta\bar{n}=0.0014n_\mathrm{p}$ using a pre-modulated plasma, and the result is shown in Fig. \ref{phase_velocity}(a). When the modulation period is much longer than the electrons' excursion, the $\delta \bar{v}_\phi$ extracted from simulations agrees well with the formula. A large deviation between the formula and the simulation results is found when $\lambda_\mathrm{m} \lesssim c/\omega_\mathrm{p}$ for this set of parameters. When $\lambda_\mathrm{m} \lesssim 0.24~c/\omega_\mathrm{p}~(400~\nano\meter)$, $\delta \bar{v}_\phi$ starts to decrease as $\lambda_\mathrm{m}$ decreases. Specifically, when $\lambda_\mathrm{m}=0.12c/\omega_\mathrm{p}~(200~\nano\meter)$, $\delta \bar{v}_\phi$ is one order of magnitude smaller than the formula. We also show $\delta \bar{v}_\phi$ from a counter-propagating laser-modulated simulation (red cross) in Fig. \ref{phase_velocity}(a) which is slightly smaller than that from the pre-modulated case. Note that the ion density is uniform in the laser-modulated case and the ponderomotive force of the standing wave plays the role of the ion density modulation for a pre-modulated plasma \cite{xu2022generation}. In Fig. \ref{phase_velocity}(b), we also plot the electron oscillation excursion amplitude v.s. $a_0$ while fixing $\lambda_\mathrm{m}=0.24~c/\omega_\mathrm{p}$. It is found that the excursion scales as $z_\mathrm{osc} \propto a_0^2$, and $\delta\bar{v}_\phi$ approaches the theoretical predication $0.12c$ as $a_0$ decreases. 

\section{All-optical pre-bunched beam generation} \label{section3}
We next carry out simulations with all the elements for the scheme included self-consistently. The quasi-3D version of OSIRIS is used with a customized Maxwell solver \cite{li2017controlling, xu2020numerical} which can eliminate numerical Cherenkov radiation \cite{godfrey1974numerical, xu2013numerical} and nonphysical space-charge fields \cite{xu2020numerical} to model the motion of the injected electrons with high fidelity. A linearly polarized (LP) laser pulse driver with $a_0=4$, a spot size of $w_0=4k_\mathrm{p}^{-1}\approx 6.7~\micro\meter$ and a FWHM intensity duration of 25 fs (4.46 $\omega_\mathrm{p}^{-1}$) propagates into the plasma 280 fs after the head of the co-propagating colliding laser and excites a fully nonlinear plasma wake surrounded by a high-density electron sheath \cite{PhysRevLett.96.165002} [Fig.\ref{schematic_plot}(a)]. The polarization of the drive laser is perpendicular to that of the colliding lasers to prevent the drive laser from contributing to the density modulation. The wavelength (frequency) of the drive laser and the colliding lasers can be different, however for simplicity we choose a wavelength of $800~\nano\meter$ for all three pulses. The plasma density is $n_\mathrm{p}=10^{19}~\centi\meter^{-3}$ which corresponds to $\omega_0/\omega_\mathrm{p}\approx 13.2$ or $n_\mathrm{p}/n_\mathrm{c}\approx 0.0057$, where $n_\mathrm{c}$ is the critical density corresponding to 800 nm wavelength. The drive laser has a peak power of 24 TW, corresponding to $P/P_\mathrm{c}\equiv \frac{a_0^2 k_\mathrm{p}^2w_0^2}{32}=8$ where $P_\mathrm{c}$ is the critical power for self-focusing \cite{Sprangle87} ($P_\mathrm{c}~[\giga\watt]\approx 17 \frac{\omega_0^2}{\omega_\mathrm{p}^2}$). It is not possible to perfectly match the laser, therefore for the chosen pulse length the projected spot size increases (laser diffracts) initially and it is subsequently self-focused by the plasma \cite{XuPRAB2023}. We first show results from a simulation where only the drive laser is present. Figure \ref{injection}(a) shows the evolution of the peak $a_0$ and the motion of the center of the wake (which is defined as the position where $E_z=0$) as a function of the location of the front of the laser, $z$. The velocity of the wake center $v_{\phi, E_z=0}$ is lower than $c$ for $z \approx 45~\micro\meter$ and $z$ between $65~\micro\meter$ and $145~\micro\meter$. Note that the phase velocity of the tail of the wake is what determines injection, however diagnosing the phase velocity of the tail is complicated by the presence of the injected electrons. Thus we use $v_\phi$ of the wake center to infer $v_\phi$ of the tail, i.e., we assume $(c-v_{\phi})/c\approx  \left(\mathrm{d}\omega_\mathrm{p}/\mathrm{d}z\right)\omega_\mathrm{p}^{-1}\xi$ so that the value of $c-v_\phi$ at the tail is approximately twice that of the center.

The sheath electrons can obtain a large forward velocity when they return towards the axis at the tail of the wake and some of them acquire a speed faster than the phase velocity of the wake's tail. They can thus become trapped during the wake expansion \cite{xu2017high,XuPRAB2023}. The plasma downramp is absent in this setup and the evolution of the laser pulse plays a similar role as the downramp which is to expand the wake and inject electrons. 

Figure \ref{injection}(b) shows the charge density distribution of the injected electrons in the ($z_\mathrm{i}, \xi$) space when the colliding lasers are not present, where $z_\mathrm{i}$ is the initial longitudinal position of the electrons and $\xi$ is their position in the speed of light frame after injection relative to the front of the drive laser. There is an approximate linear mapping between $z_\mathrm{i}$ and $\xi$ for electrons \cite{xu2017high} with $z_\mathrm{i} \gtrsim 60~\micro\meter$. This mapping originates from a continuous expansion of the wake and ensures the generation of a pre-bunched beam if the injection is turned on and off frequently by the modulated plasma density. Due to the initial evolution of the laser, the injection before $z_\mathrm{i}\approx 60~\micro\meter$ is not as organized. This figure also shows that the injected beam duration $\delta \xi \sim \micro\meter$ is significantly compressed when compared to the injection duration $\delta z_\mathrm{i} \sim 100~\micro\meter$. 

When the drive laser and the co-propagating laser collide with the counter-propagating laser, each location ($\xi$) with in the drive laser will now encounter a plasma with a density modulation at $2\omega_\mathrm{1}$. The spot size of colliding lasers are chosen to be significantly larger than that of the drive laser (10 $\micro\meter$ compared to 6.7 $\micro\meter$), to ensure that all radii of the wake experience considerable density modulation. The phase velocity of the wake's tail is therefore additionally modulated at a frequency of $2\omega_\mathrm{1}$. If the amplitude of this modulation is large enough then the injection can be turned on and off at a frequency of $2\omega_\mathrm{1}$, forming a pre-bunched beam with a period of $ 7~\nano\meter$. The beam loading effect causes a slow reduction of the $v_\phi$ at the tail, which does not disrupt the high frequency discrete injection and pre-bunched beam generation, as long as the modulation amplitude of $v_\phi$ is properly chosen.

The longitudinal phase space (blue) and the current profile (red) of the injected beam are shown in Fig. \ref{injection}(c). An initial energy chirp of $\sim 80~\mega\electronvolt/\micro\meter$ is present because electrons at the head of the beam have been accelerated for longer times than those at the tail. This energy chirp can be subsequently canceled by the negatively chirped wakefield [see Fig. \ref{pol}(a)] over some acceleration distance \cite{xu2022generation}. The current profile shows a clear but small amplitude modulation for $\xi \lesssim 21.4~\micro\meter$. The inset provides a zoomed in view of the current profile which clearly shows the periodic modulation. The main body of the injected beam has a small slice normalized emittance of $\sim 30~\nano\meter$ and a slice energy spread of $\sim0.5~\mega\electronvolt$ for this part \cite{XuPRAB2023}. On the other hand, the head of the high current beam is characterized by a relatively large normalized emittance ($>100 ~\nano\meter$) and a large energy spread (several MeV). The bright, chirped and pre-bunched main body of the beam can enable a self-selection FEL process to produce attosecond radiation when colliding with an optical undulator at the plasma exit \cite{xu2023atto}. If the beam is further accelerated to GeV energies with low energy spread, then a seeded XFEL in a magnetic undulator is possible \cite{xu2022generation}. The slice energy spread of the injected beam may degrade the pre-bunched structure. The high acceleration gradient of the nonlinear plasma wake can boost the beam energy rapidly and suppress the degradation during acceleration. After the plasma, the slice energy spread poses a limitation on the beam drift distance as $L_\mathrm{d}\ll \frac{\gamma_\mathrm{b}^3}{\sigma_\gamma}\lambda_\mathrm{b}$, where $\gamma_\mathrm{b}= \frac{E_\mathrm{b}}{mc^2}$ is the relativistic factor of the beam upon exiting the plasma, $\sigma_\gamma = \frac{\sigma_\mathrm{E}}{mc^2}$ is the slice energy spread, and $\lambda_\mathrm{b}$ is the pre-bunched wavelength.

The bunching factor, $b(k)=\left| \sum_{j=1}^N \mathrm{exp}(ikz_j)\right|/N$, is widely used to quantify the modulation of pre-bunched beams, where $N$ is the number of the electrons. We show the bunching factor for electrons with $\xi < 21.4~\micro\meter$ in Fig. \ref{injection}(d). There is a peak at $k\approx 115 k_\mathrm{1}$ ($\lambda\approx 7~\nano\meter$) with $b\approx 0.01$. The modulation of the current profile is shallow and we do not observe a distinct energy modulation caused by the beam-plasma oscillation \cite{xu2022generation}. The harmonic number $h$ which is defined as the ratio between the wavelength of the colliding lasers and the pre-bunching period is $h\approx 115$. Since the electrons are injected at the tail of the nonlinear wake, the harmonic number can be written as $h\approx \frac{\Delta z_\mathrm{i}}{\Delta \xi} \approx \frac{\Delta z_\mathrm{i}}{(c - v_\phi)(\Delta z_\mathrm{i}/c)}\approx 2\gamma_\phi^2$, which has the same $\gamma_\phi$ dependence as in a relativistic Doppler shift. A wake with a slower expansion rate results in a larger $v_\phi$ and a larger $h$, and vice versa. 

The phase velocity modulation induced by the colliding lasers with $a_\mathrm{1}=0.002$ (corresponding to $\delta \bar{n} / n_\mathrm{p}\approx 1.4\times 10^{-3}$) is sufficiently large to turn the injection on and off. We scan $\delta \bar{n} / n_\mathrm{p}$ from $2.5\times 10^{-4}$ to $4\times 10^{-3}$ in a pre-modulated plasma and find the bunching factor remains similar. The modulation depth of the injected current is dominated by the spread of $\xi$ for electrons originating from the same $z_\mathrm{i}$. We show the $\xi$ distribution for two representative $z_\mathrm{i}$ in Fig. \ref{injection}(b). For $60~\micro\meter < z_\mathrm{i} < 80 \micro\meter$, the FWHM width of $\xi$ is $5\sim 10~\nano\meter$. There is a split of the mapping for electrons with $z_\mathrm{i}\gtrsim 80~\micro\meter$, i.e., the $\xi$-distribution at a fixed $z$ exhibits a two-peak or three-peak structure. The width of the main peak is even shorter than that between $60~\micro\meter < z_\mathrm{i} < 80~\micro\meter$. Figures \ref{injection}(e) and (f) show the plasma electron density distribution at two different propagation distances. The sheath surrounding the wake has one layer at $z=67.3~\micro\meter$ while it splits into 7 layers at $z=101~\micro\meter$ \cite{10.1063/5.0051282} and only the electrons in the innermost 2 or 3 layers can obtain enough forward velocity to be trapped. We can see the split of the mapping is consistent with the sheath structure of the nonlinear wake. This multi-layer fine structure will be smoothed out by a finite plasma temperature but the larger ones remain.

The pre-bunched wavelength can be decreased by using colliding lasers with shorter wavelengths \cite{xu2022generation}. However, the minimum period is limited by the aforementioned $\xi$-spread in $\xi$. Simulations using a pre-modulated plasma indicate that when $\lambda_\mathrm{m}=300~\nano\meter$ (corresponding to a $5.3~\nano\meter$ pre-bunching wavelength), the bunching factor decreases to only one-fourth of its value with $\lambda_\mathrm{m}=400~\nano\meter$. Note that a pre-modulated plasma with a period of $\lambda_\mathrm{m}$ is equivalent to a laser-modulated plasma with $\lambda_1=2\lambda_\mathrm{m}$. It is possible to fine-tune the parameters of the drive laser to excite nonlinear wakes with narrow and simple sheath structures. In general, the wakes excited by tightly focused beam drivers exhibit a single-layer and narrow sheath compared to those produced by laser drivers. Thus, an electron beam driver can generate pre-bunched beams with deeper modulation and shorter periods, at the expense of a more complicated setup.

\section{Effects of the colliding laser pulses on the pre-bunched beams}
\label{section4}

\begin{figure*}[htbp]
\centering
\includegraphics[width=0.7\linewidth]{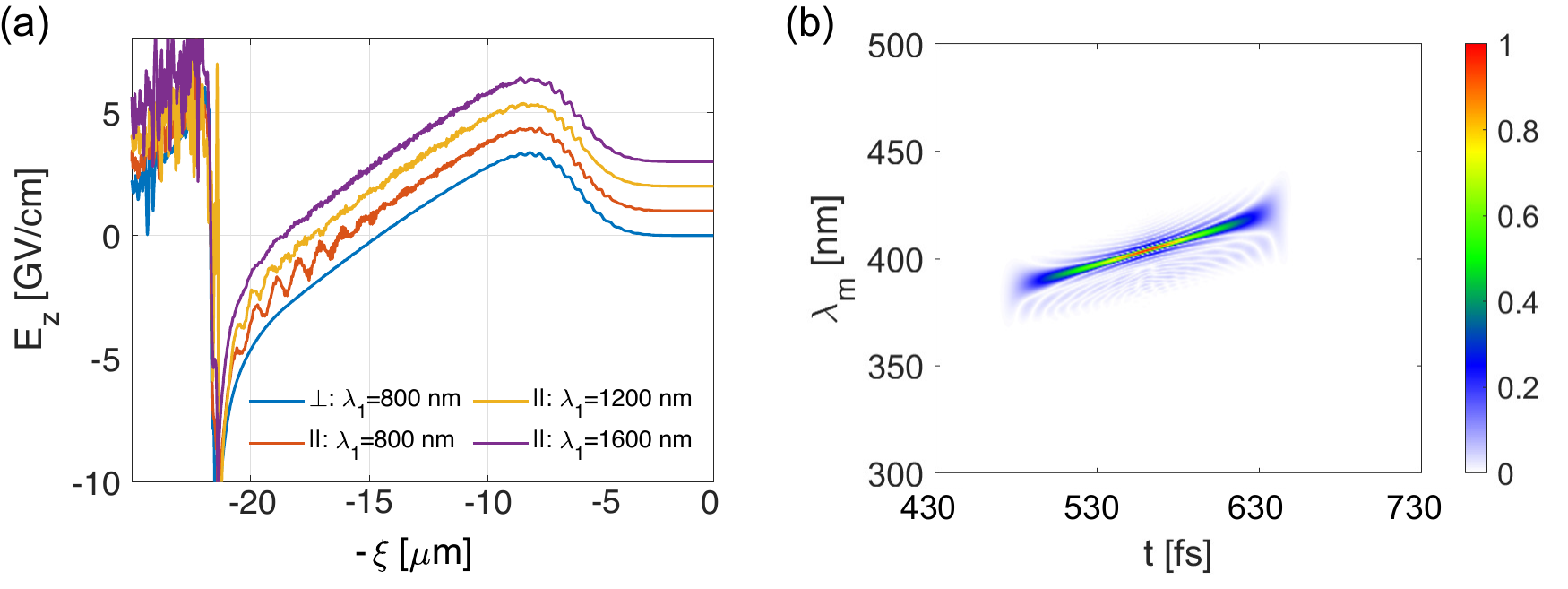}
\caption{The effects of the colliding lasers on the pre-bunched beams. (a) On-axis $E_z$ of the plasma wake driven by a 800 nm wavelength drive laser for different polarization and wavelengths of the colliding pulses. The lines are vertically offset. (b) The Wigner function of the density modulation during the injection when two colliding pulses with $\beta = 6.3~\femto\second^{-2}$ are used. }\label{pol}
\end{figure*}

The parameters of the colliding lasers, such as envelope profile, duration, intensity, frequency, and polarization, can affect the properties of the injected electrons. In this section, we examine the effects of a few of these parameters. We first consider what happens if the colliding lasers have two distinct frequencies, $\omega_\mathrm{r}$ and $\omega_\mathrm{l}$, where `r' refers to right going and `l' refers to left going. If the vector potentials for each wave have equal amplitude then the total normalized vector potential is $a=a_1 \left[ \mathrm{cos}(k_\mathrm{r}z-\omega_\mathrm{r}t) + \mathrm{cos}(k_\mathrm{l}z+\omega_\mathrm{l}t) \right]=2 a_1 \mathrm{cos}\left(  \bar{k}z - \Delta \omega t \right) \mathrm{cos}\left( \Delta k z - \bar{\omega} t\right)$, where $\bar{k}= \frac{k_\mathrm{r}+k_\mathrm{l}}{2},\bar{\omega}= \frac{\omega_\mathrm{r}+\omega_\mathrm{l}}{2},\Delta k=\frac{k_\mathrm{r}-k_\mathrm{l}}{2},\Delta \omega=\frac{\omega_\mathrm{r}-\omega_\mathrm{l}}{2}$. The ponderomotive force is $F_\mathrm{p}\propto \langle a^2 \rangle$, where $\langle \rangle$ refers to averaging over the high frequencies. Upon making the substitutions and assuming $\Delta \omega \ll \bar{\omega}$, we obtain $F_\mathrm{p}\propto \mathrm{cos}\left( 2\bar{k}z - 2\Delta \omega t\right) \approx \mathrm{cos}\left( k_m z - 2\frac{\Delta \omega}{c} \xi\right) $, where $k_\mathrm{m}=k_\mathrm{r}+k_\mathrm{l}$ and Eq. \ref{deltavphi} is modified to $\delta v_\phi \approx  -c\frac{\delta \bar{n}}{n_\mathrm{p}}  \frac{k_\mathrm{m}}{2}\xi \mathrm{sin}(k_\mathrm{m}z - 2\frac{\Delta \omega}{c}\xi)$. For slow expansions of the bubble and injection inside the first bucket, the $2\frac{\Delta \omega}{c}\xi$ term can be neglected. If $a_\mathrm{1r}$ and $a_\mathrm{1l}$ are not equal then the $a_1^2$ factor in the expression of $\delta \bar{n}$ should be replaced by $\left( \frac{a_\mathrm{1r} + a_\mathrm{1l} }{2} \right)^2$.

The colliding pulses must be linearly polarized along the same direction to form a standing wave and for the simulations used for the case in Fig. \ref{injection} the polarization direction of the drive laser is chosen to be perpendicular to that of the colliding lasers. If they are parallel, the beating between the laser driver and the counter-propagating colliding lasers provides a ponderomotive force that modifies the plasma wake and the injection dynamics. Quasi-3D simulation results are shown in Fig. \ref{pol} to illustrate the importance of the relative polarization. A well behaved wake is formed with perpendicular polarizations while the second half of the wake is modulated significantly with parallel polarizations which destroys the subtle injection dynamics. A quasi-3D simulation confirms that a pre-bunched structure is absent for parallel polarizations as shown in Fig. \ref{injection}(d) (red curve). This also indicates that it is not feasible to pre-bunch the injected electrons using only two pulses with the same frequencies: the drive laser and the counter-propagating laser. 

On the other hand, the beating between the ponderomotive forces of the drive laser and the counter-propagating colliding laser that arise for parallel polarization can be eliminated by introducing a large frequency (wavelength) difference between them. As shown in Fig. \ref{pol}(a), the modulation ripples on the $E_z$ field disappear when $\lambda_\mathrm{1}=1600~\nano\meter$ and are reduced for $\lambda_\mathrm{1}=1200~\nano\meter$. In the 1600 nm case, a pre-bunched beam with a period of $16~\nano\meter$ and $b=0.025$ is injected as shown in Fig. \ref{injection}(d).

We can vary the duration of the colliding pulses and/or the delay between the driver and the colliding pulses to limit the extent of the modulated plasma region encountered by the driver, thereby producing a beam that is partially pre-modulated. When a beam with a pre-modulated tail propagates inside a resonant undulator, the superradiant radiation emitted from the tail can extract energy from ``fresh" electrons efficiently when slipping forward, thus producing radiation pulses whose power grows as the propagation distance squared ($P\propto z^2$), and whose duration decreases as the inverse square root of the distance ($\tau \propto z^{-1/2}$) \cite{franz2024terawatt}. Thus, by sending beams with pre-bunched tails into an optical undulator, we may generate high-power attosecond x-ray pulses in an all-optical and extremely compact manner. 

Furthermore, we can replace the long backward propagating colliding laser pulse by a series of short pulses with controllable durations and separations to pre-modulate the injected electrons in a series of positions. The use of lasers with flying foci could add even more flexibility and control \cite{froula2018spatiotemporal}. These beams can produce a train of attosecond radiation pulses in a resonant undulator. If backward short pulses have fixed relative phases and are delayed equally, these properties can be inherited by the density modulation of the electron beam which can enable mode-locked FELs \cite{PhysRevLett.100.203901} to produce ultrashort radiation pulse trains with a comb-like spectrum. 

It is also feasible to produce exotic plasma density modulation and precisely map them to the injected beams. If two colliding lasers with a linear frequency chirp $\beta=\frac{1}{2}\frac{\mathrm{d}\omega}{\mathrm{d}t}$ are used, the driver can encounter a density modulation with a chirped wavelength over the injection distance. An example is shown in Fig. \ref{pol}(b) which represents the Wigner function of density modulation during the injection, where colliding lasers with $\beta=3.2\times 10^{-4}~\femto\second^{-2}$ are used. Experiments conducted at an XFEL facility have shown an electron beam with chirped pre-bunched period can emit chirped radiation in the extreme-ultraviolet spectrum in an undulator which can be further compressed to sub-femtosecond duration and GW power \cite{gauthier2016chirped}. Our all-optical pre-bunched beam generation concept might push this scheme into the soft x-ray spectrum to generate ultrashort and high-power radiation. Furthermore, an electron beam with a chirped pre-bunched structure can enable the generation of an isolated monocycle x-ray pulse in a tapered undulator \cite{PhysRevLett.114.044801}.

\section{Conclusion}
In summary, we have proposed an all-optical PBA-based scheme that can generate high brightness, low energy spread, multi 10 kA electron beams that are pre-bunched at nanometer scales. Such beams can be used to drive compact XFELs. An intense drive laser creates an expanding wakefield thereby slowing down its phase velocity to trigger self-injection. When the drive laser also encounters density modulations caused by two low intensity colliding lasers, the wakefield's phase velocity is also modulated such that the injection is turned on and off periodically. As a result of the longitudinal mapping between the injection time and the phase in the wakefield, a pre-bunched beam is formed. Additionally, by controlling the properties of the colliding lasers, the plasma density modulation can be tuned to produce injected beams with chirped bunching period or partially bunched beams. These exotic beams can enable novel FEL operation modes and generate ultrafast high power x-rays. This all-optical and extremely compact pre-bunched beam generation scheme may bring more opportunities to next-generation compact x-ray sources and enhance the capabilities at existing x-ray user facilities. 

This concept also has much flexibility. The properties of the drive laser affect the injection process and the properties of the injected beam, such as the injection region, the brightness, the pulse duration and the compression ratio. This was discussed thoroughly in previous publications \cite{XuPRAB2023}. Additionally, a density ramp with a small density gradient can be utilized to further control the harmonic number, i.e., a density upramp (downramp) can increase (decrease) $h$. For instance, $h$ is decreased to $58$ in a downramp ranging from $z=63~\micro\meter$ to $105~\micro\meter$ with a density jump of $0.1n_\mathrm{p}$. The proposed scheme has a loose requirement on the synchronization between these three laser pulses, i.e., the driver laser needs to experience a density-modulated plasma during the injection. Since hundreds of fs long colliding pulses are used and their induced plasma density modulation can sustain hundreds of fs, the synchronization between these three laser pulses is $\sim 100$ fs which is readily realized in contemporary high-power laser facilities.

\begin{acknowledgments}
This work was supported by the National Natural Science Foundation of China (Grant No. 12375147, No. 11921006), Beijing outstanding young scientist project, National Grand Instrument Project No. SQ2019YFF01014400, and the Fundamental Research Funds for the Central Universities, Peking University. The simulations were supported by the High-performance Computing Platform of Peking University and Tianhe new generation supercomputer at National Supercomputer Center in Tianjin.
\end{acknowledgments}

\section{data availability}
 The data that support the findings of this article are available from the corresponding author upon the reasonable request.

\appendix
\section{Setup of Particle-in-cell Simulations}
For the fully self-consistent quasi-3D simulations of the all-optical scheme to generate pre-bunched injected beams, we use a fixed window with a box size of $125k_\mathrm{p}^{-1}\times 12k_\mathrm{p}^{-1}$ and $100000\times 2400$ cells along the $z$ and $x$ directions, respectively. The plasma rises smoothly from $z=0$ to $10^{19}~\centi\meter^{-3}$ at $z=10k_\mathrm{p}^{-1}$ to eliminate possible injection from the boundary and then remains constant until $z=100k_\mathrm{p}^{-1}$ where drops sharply to 0. The grid size along the $z$ direction is $\mathrm{d}z=\frac{1}{800}k_\mathrm{p}^{-1}$ to ensure there are $\sim4$ points per bunching wavelength and the transverse grid size is $\mathrm{d}x=\frac{1}{200}k_\mathrm{p}^{-1}$ to resolve the thin wake sheath structure. 32 macro-particles per cell are used to represent the pre-ionized plasma electrons. The rising and falling edges of the colliding lasers and the drive laser have a polynomial profile of $10\tau^3 - 15\tau^4 + 6\tau^5$, where $\tau=\frac{t}{t_\mathrm{rise}}$ or $\tau=\frac{t}{t_\mathrm{fall}}$. At the start of the simulation ($t=-12.5\omega_\mathrm{p}^{-1}$), the leading edge of the colliding laser pulses are located at the left and right edge of the simulation window, i.e., $z=-12.5k_\mathrm{p}^{-1}$ and $112.5k_\mathrm{p}^{-1}$. The leading edge of the drive laser is located at $z=0$ at the start of the simulation and it is delayed by 62.5 $\omega_\mathrm{p}^{-1}$. The driver laser is symmetric with $t_\mathrm{rise}=t_\mathrm{fall}=5.8~\omega_\mathrm{p}^{-1}$. Recall $\omega_\mathrm{p}^{-1}=5.6 ~\femto\second$ and $c/\omega_\mathrm{p}=1.68~\micro\meter$. Recently developed advanced Maxwell solvers \cite{xu2020numerical, li2021new} are needed for modelling the sutble injection and acceleration of the bright pre-bunched beams. 

A moving window propagating at the speed of light in vacuum $c$ is used in simulations without colliding laser pulses. The box size is $14k_\mathrm{p}^{-1}\times 12k_\mathrm{p}^{-1}$ and there are $11200\times 2400$ cells along the $z$ and $x$ directions, respectively. Other settings are the same as the simulations with colliding lasers.

The 1D simulations with the colliding pulses use a fixed window with a box size of $125k_\mathrm{p}^{-1}$ and $20000$ cells along the $z$ direction, and 128 macro-particles per cell to represent the plasma electrons. A moving window propagating at $c$ with a size of $20k_\mathrm{p}^{-1}$ and 64000 grids is used in the 1D simulations without colliding pulses. The same plasma profile as in the quasi-3D simulations is used.

\normalem\bibliography{refs_xinlu}% Produces the bibliography via BibTeX.

\end{document}